\documentstyle{article}

\def\Tr{\mathop{\rm Tr}\nolimits}
\def\arccot{\mathop{\rm arccot}\nolimits}
\def\intm#1{\int {d^2 #1 \over (2\pi)^2} \,}

\def\intk{\intm{k}}
\newcommand{\half}{{\textstyle\frac12}}

\title{\rightline{\tenrm IFUP-TH 49/92}\bigskip
$1/N$ Expansion of Two-Dimensional Models in the Scaling Region
\footnote{Talk presented at the {\em Lattice '92\/} Conference, Amsterdam.}
}

\author{%
Massimo Campostrini and Paolo Rossi\\
\tenit Istituto Nazionale di Fisica Nucleare, Sezione di Pisa,\\
\tenit Dipartimento di Fisica dell'Universit\`a, I-56126 Pisa, Italy.%
}

\date{October 6th, 1992}

\begin{document}

\maketitle

\begin{abstract}
The main technical and conceptual features of the lattice $1/N$
expansion in the scaling region are discussed in the context of a
two-parameter two-dimensional spin model interpolating between $CP^{N-1}$
and $O(2N)$ $\sigma$ models, with standard and improved lattice actions.

We show how to perform the asymptotic expansion of effective propagators
for small values of the mass gap and how to employ this result in the
evaluation of physical quantities in the scaling regime.

The lattice renormalization group $\beta$ function is constructed
explicitly and exactly to $O({1/N})$.
\end{abstract}

\section{Introduction}
Lattice field theories are the natural ground of application of numerical
techniques.

However the numerical results are usually affected by two major
limitations:

the very slow approach to the critical (scaling) domain when we compare the
quality of the results to the numerical effort;

the lack of control on the reliability of some of the lattice definitions
and methods: we only mention here the problems arising in the definition of
the topological charges, in the subtraction of the perturbative tails, in
the extrapolation to the chiral limit, in the checks of ergodicity.

An analytical approach offers the possibility of testing methods and
results of numerical simulations in a controlled environment, and it can
therefore be illuminating even if its domain of direct applicability is
rather restricted.

In particular the $1/N$ expansion of two-dimensional spin models has
the following relevant advantages:

the expansion parameter is not subject to renormalization and does not
depend on the physical scale; the resulting series presumably has a finite
convergence radius;

results are RG invariant; dynamically generated masses appear naturally in
the computation and Borel ambiguities in the resummations are bypassed;

the $1/N$ expansion of spin models is technically viable and in two
dimensions it may deal with nontrivial asymptotically free theories and
interesting topological structure;

scaling behaviours on the lattice may be explicitly analyzed;

numerical simulations are easily available.

\section{The Model}
As an illustration of our techniques we consider a model described by the
continuum action \cite{Samuel}
\begin{equation}
S = N \int d^2 x \left\{\beta_{\rm v}\,\partial_\mu\bar z\partial_\mu z +
\beta_{\rm g}\,\overline{D_\mu z}D_\mu z \right\},
\end{equation}
where $\bar z z = 1$ and $D_{\mu} = \partial_{\mu} + i \bar z
\partial_{\mu} z$,
interpolating between $CP^{N-1}$ $(\beta_{\rm v} = 0)$ and $O(2N)$
$(\beta_{\rm g} = 0)$.

We can introduce the effective couplings
${1 \over2 f} = {\beta_{\rm v} + \beta_{\rm g}}$
and
$ {\kappa} = 2 (\beta_{\rm v} + \beta_{\rm g}) {\beta_{\rm v}
\over{\beta_{\rm g}}}$

The coupling $f$ is running and asymptotically free, while $\kappa$ does not
renormalize and lines of constant $\kappa$ are RG trajectories.

Integrating over the fundamental fields $z$ after the introduction of a
scalar and a vector multiplier field in order to implement the constraints
we obtain the effective action
\begin{eqnarray}
\lefteqn{S_{\rm eff} = N \Tr\ln\left\{-\Box
- i\left\{\partial_\mu,\theta_\mu\right\} + \alpha\right\}} \nonumber \\
&&+ {N\over2f}\left\{-i\alpha + (1+\kappa f)\theta_\mu\theta_\mu\right\}.
\end{eqnarray}

The phase diagram can be briefly described as follows:

At $\kappa = 0$ a linear confining potential leads to the disappearance
of the $z$ fields from the physical spectrum. A pure $\bar z z$ bound state
spectrum is observed \cite{DiVecchia,Witten}.

At small $\kappa$ deconfinement occurs, but the fundamental fields are
still very heavy compared to the bound states.

At larger $\kappa$ a crossover occurs, until at $\kappa = {1/\pi}$
bound states disappear from the physical spectrum.
When  $\kappa = {1/\pi}$  a conserved nonlocal quantum charge exists:
scattering is factorized and an exact S matrix can be reconstructed
\cite{Abdalla}.

When $\kappa = {n/\pi}$, $n$ integer, we are dealing with an effective
gauge-fixed version of $CP^{N-1}$ with n flavours of minimally coupled
massless fermions.

At $\kappa = \infty$ the $U(N)$ symmetry is promoted to $O(2N)$.

The dynamical mass generation is governed by the gap equation
\begin{equation}
{1\over2f} = \intk {1\over k^2 + m_0^2} \,,
\end{equation}
implying
\begin{equation}
m_0^2 \sim \exp\left(-{\pi\over f}\right).
\end{equation}

The singularity of the vector propagator occurs at $\mu$ defined in the
large N limit by
\begin{equation}
\sqrt{{4m_0^2\over\mu^2}-1}\;
\arccot\sqrt{{4m_0^2\over\mu^2}-1} = 1-\pi\kappa.
\end{equation}

The $1/N$ expansion allows us to compute explicitly the $O({1/N})$
correction to the location of the mass pole
\begin{equation}
m_1^2 = \Sigma_1\left(-m_0^2\right),
\end{equation}
where $\Sigma_1$ is the $O({1/N})$ contribution to the self-energy
function.

We can also extract the scaling part of the free energy and find the
(scheme-independent) relationship
\begin{eqnarray}
\lefteqn{{F\over m^2} = {1\over4\pi}\left[N + C(k)
 + O\left(1\over N\right)\right],} \nonumber \\
\lefteqn{C\left(1\over\pi\right) = 0, \quad C(\infty) = -1.}
\end{eqnarray}

\section{Lattice Formulation and Asymptotic Expansion}
In the lattice formulation we introduce a nearest-neighbor lattice action
\begin{eqnarray}
\lefteqn{S = N\sum_{n,\mu}\left\{ \beta_{\rm v}\, \bar z_{n+\mu} z_n +
\beta_{\rm g}\, \bar z_{n+\mu}\lambda_{n,\mu}z_n + \hbox{h.c.}\right\},}
\nonumber \\
\end{eqnarray}
where $\lambda_{\mu} = \exp(i \theta_{\mu})$,
and its Symanzik tree-improved counterpart, involving second-nearest
interactions.

Integrating over the z fields we obtain the lattice effective action as a
function of $\alpha$ and ${\theta_\mu}$.

Taking functional derivatives with respect to these fields we obtain
effective propagators $\Delta$ and effective vertices, both amenable to
standard
one-loop lattice integrals.

The $1/N$ expansion is a loop expansion in the effective propagator
loops.

The study of the scaling region requires expanding lattice quantities in
powers of $m^2_0$; in particular \cite{Biscari}
\begin{eqnarray}
\lefteqn{\Delta^{-1}(k) = \sum_{n=0}^\infty
\left[A_n(k)+\beta B_n(k)\right] m_0^{2n},} \nonumber \\
&&{\beta = {1\over2f}\sim \ln m_0^2.}
\end{eqnarray}
These expansions are asymptotic; the introduction of an explicit dependence
on $\beta$ intead of $\log(m^2_0)$ allows for a direct comparison with
standard perturbation theory.

$B_n(k)$ can always be computed in closed form.
$A_n(k)$ have regular integral representations.

The lattice evaluation of physical quantities in the scaling region should
not in principle lead to divergent integrals. However IR divergences may
appear at intermediate stages of the computation, due to the asymptotic
character of the expansion. One can regroup the contributions in such a way
that they naturally decompose into two classes:

IR regulated massless lattice integrals;

UV regulated massive continuum integrals, corresponding to a specific
continuum regularization (``sharp momentum cutoff'': SM)

It is easy to show that adimensional ratios of physical quantities are
universal, i.e. the contribution from the massless lattice integrals
cancels out.

It is therefore sufficient to compute a single dimensionful quantity on the
lattice, e.g. the mass gap, in order to make sensible predictions for all
physical quantities (with canonical scaling dimensions) in the lattice
scaling region.

Assuming the quantity $Q$ has canonical dimension $2 \gamma$ and its
continuum (SM) value is
\begin{equation}
Q = (m_0^2)^\gamma\left[q_0 + {1\over N}q_1(\beta)
+ O\left(1\over N^2\right)\right]
\end{equation}
on the lattice we obtain
\begin{eqnarray}
\lefteqn{Q^{(2)} = (m_0^2)^\gamma\biggl[q_0 + {1\over N}q_1(\beta)
+ {1\over N}\gamma q_0 {\delta m_1^2(\beta)\over m_0^2}} \nonumber \\
&&+ O\left(m_0^2\right) + O\left(1\over N^2\right)\biggr],
\end{eqnarray}
where
\begin{equation}
\delta m_1^2 = m_{1\,\rm L}^2 - m_{1\,\rm SM}^2.
\end{equation}

We can also prove that
\begin{eqnarray}
\lefteqn{\left[m_{1\,\rm L}^2\right]_{\rm scal} =
\half\Delta_{(\alpha)}(0) \intk}\nonumber \\
&&\times{\partial\over\partial m_0^2}\left[\ln\Delta^{-1}_{(\alpha)}(k)
+ \ln\det\Delta^{-1}_{\mu\nu}(k)\right]
\end{eqnarray}
plus continuum (SM) counterterms.

In general we may therefore extract the $O({1/N})$ lattice RG $\beta$
function. For nearest neighbor interactions we obtained
\begin{eqnarray}
\lefteqn{\beta_1^{({\rm L})}(f) = \beta_1^{({\rm SM})}(f)
+ {f^2\over\pi N}\,{\beta^2\over2\pi}\,
  {\partial\over\partial f}{\delta m_1^2\over m_0^2}} \nonumber \\
&&= -{f^2\over\pi}\Biggl\{1 + {1\over N}\,{f\over2\pi}
  \left(1 + {3-2\pi\tilde\kappa \over
        1 + f\left(\tilde\kappa-1/\pi\right)}\right) \nonumber \\
&&-{1\over2N}\intk{A_1^\alpha B_0^\alpha - B_1^\alpha A_0^\alpha \over
                   \left(A_0 + \beta B_0\right)^2} \nonumber \\
&&-{1\over2N}\intk\Biggl[{A_1^\theta B_0^\theta - B_1^\theta A_0^\theta
        \over \left(A_0 + \beta B_0\right)^2} \nonumber \\
&&\quad + {\tilde\kappa\left[\left(A_0^\theta + 4\beta^2B_0^\theta\right)
                 - (4\beta-1)^2B_1^\theta\right] + {\tilde\kappa}^2 \over
        (4\beta-1)^2\left(A_0 + \beta B_0\right)^2} \Biggr] \nonumber \\
&&-{1\over2N}\,{1\over(4\beta-1)^2}
  +{1\over N}\int^{\sqrt{32}} {d^2 k \over (2\pi)^2}\,{4\pi\over k^2}
                \nonumber \\
&&\times \left[{1\over\ln\left(k^2/m_0^2\right)} +
        {3-2\pi\tilde\kappa \over\left(\ln\left(k^2/m_0^2\right)
                + 2\pi\tilde\kappa - 2\right)}\right]\Biggr\} \nonumber \\
\end{eqnarray}
where
\begin{equation}
\tilde\kappa = \kappa\left(1-{1\over4\beta}\right).
\end{equation}

Integrals are series-expandable in powers of $f$ reproducing standard
perturbation theory to $O({1/N})$ and to all orders in $f$.
The integrals can also be computed by a principal-part prescription and
thus offer a nonambiguous resummation of the perturbative series \cite{noi}.

\end{document}